# PLASMA BREMSSTRAHLUNG EMISSION AT ELECTRON ENERGY FROM LOW UP TO EXTREME RELATIVISTIC VALUES

## Alexei Yu. Chirkov

Bauman Moscow State Technical University, Moscow, Russia,
e-mail: alexxeich@mail.ru

**Abstract** – We consider the bremsstrahlung of electrons of fully ionized plasma. Electron energy range extends from a few hundred electron volts before the extreme relativistic energies of several MeV. The results of calculations of effective slow-down rate of a single electron and the power emission from a volume of the plasma with Maxwellian electrons. At relativistic energies, the Born approximation yields highly accurate results. Correction to the Born approximation are considered for low-energy regimes. In the high-energy regimes numerical results are close to extreme relativistic analytic results. The formulas are suggested that approximate the results of numerical calculations.

Bremsstrahlung from the fully ionized plasma is considered in this paper. Here we calculate bremsstarhlung energy losses and emission power to obtain formulas for fast calculations. The results of the present study cover wider ranges of electron temperatures and energies then in previous study [1]. Temperature and energy of electrons vary from hundreds eV up to MeV in the present study. Under such conditions the plasma radiates from its volume without significant absorption even at extremely high densities available in the experiments.

Bremsstrahlung power from a unit volume is a sum of the radiation power due to electron-ion and electron-electron collisions

$$P_b = P^{ei} + P^{ee}.$$

Electron-ion bremsstrahlung power is

$$P^{ei} = \sum_i n_e n_i \iint \omega \frac{d\sigma^{ei}}{d\omega} v f(\mathbf{p}) d^3 p \, d\omega = \sum_i n_e n_i \int_0^\infty \kappa^{ei}(p) v f(p) 4\pi p^2 dp. \quad (1)$$

Here $d\sigma^{ei}$ is differential cross-section of the process; $\omega$ is the energy of the emitted photon; $\mathbf{p}$ is electron momentum; $v$ and $p$ are the absolute values of speed and momentum of an electron; $f(\mathbf{p})$ and $f(p)$ are normalized electron distribution functions;

$$\kappa^{ei}(p) = \int_0^{\omega_{\max}} \frac{d\sigma^{ei}}{d\omega} \omega d\omega$$

is the energy losses (effective slow-down rate) of an electron with initial momentum $p$;

$$\omega_{\max} = \sqrt{p^2 c^2 + (m_e c^2)^2} - m_e c^2$$

is the maximum energy of the emitted photon; $m_e$ is the rest mass of an electron; $c$ is the velocity of light.

Equilibrium Maxwellian distribution function of relativistic electron gas is [2]

$$f(p) = \frac{\exp(-\varepsilon/\theta)}{4\pi\theta(m_e c)^3 K_2(1/\theta)},$$

where $\varepsilon = 1/\sqrt{1-(v/c)^2}$ is the electron energy in units of rest energy of an electron $m_e c^2 = 511$ keV, $\theta = k_B T_e /(m_e c^2)$ is temperature of electrons expressed in units of rest energy, $K_2(...)$ is Macdonald function.

Using the expression for $d\sigma^{ei}$ in the Born approximation (relativistic Bethe–Heitler formula) [3, 4] for numerical integration we obtained the dependence of $\kappa^{ei}$ on the electron energy. Obtained data can be approximated with the error less than 1% to by the following expression:

$$\kappa^{ei} = C_b c^{-1} Z_i^2 \left\{ 4\varepsilon \left[\ln(2\varepsilon) - \tfrac{1}{3}\right] + (\tfrac{20}{3} - 4\ln 2) \exp[-0.408(\varepsilon-1)] \right\}, \qquad (2)$$

where $C_b = \alpha r_e^2 m_e c^3$, $\alpha$ is fine structure constant, $r_e$ is classical radius of the electron.

Equation (2) is constructed such a way that in the limit $\varepsilon \to 1$ it gives the non relativistic expression $\kappa_{NR}^{ei} = \tfrac{16}{3} C_b c^{-1} Z_i^2$. For $\varepsilon \gg 1$ it coincides with the extreme relativistic expression $\kappa_{ER}^{ei} = 4 C_b c^{-1} Z_i^2 \varepsilon \left[\ln(2\varepsilon) - \tfrac{1}{3}\right]$ (see Fig. 1). Using Eq. (2), we can obtain an expression for $P^{ei}$ containing a single integral

$$P^{ei} = \frac{C_b n_e^2 Z_{eff}^2}{\theta K_2(1/\theta)} \int_1^\infty \left\{ 4\varepsilon\left[\ln(2\varepsilon) - \tfrac{1}{3}\right] + (\tfrac{20}{3} - 4\ln 2)\exp[-0.408(\varepsilon-1)] \right\} \times$$
$$\times \exp(-\varepsilon/\theta)(\varepsilon^2 - 1) d\varepsilon + \Delta P^{ei}, \qquad (3)$$

where $Z_{eff}^2 = \sum_i Z_i^2 n_i \Big/ \sum_i Z_i n_i$ is the square of the effective charge; $\Delta P^{ei}$ is the correction to losses calculated using the cross section in the Born approximation.

Correction to the Born approximation is significant at temperatures $T_e < 10$ keV. For $T_e \lesssim 1$ keV Gould obtained a relative correction to Born approximation [5]

$$\delta_B = \frac{\Delta P^{ei}}{P_{NR}^{ei}} = (\ln 2 - \tfrac{1}{2}) \pi \frac{Z_{eff}^3}{Z_{eff}^2} \frac{\alpha}{\sqrt{2\theta}},$$

where

$$P_{NR}^{ei} = \frac{32}{3} \sqrt{\frac{2}{\pi}} C_b n_e^2 Z_{eff}^2 \sqrt{\theta}$$

is non relativistic expression for the power of the electron-ion bremsstrahlung concerned is the Born approximation [6], $Z_{eff}^3 = \sum_i Z_i^3 n_i \Big/ \sum_i Z_i n_i$. For $T_e = 1$ keV $\delta_B \sim 0.1$.

With increasing $T_e$ this correction decreases, while for $T_e < 10$ keV relativistic corrections are becoming more important. To remove an unlimited increase of $\delta_B$ at $T_e \to 0$, it is necessary to take into account the known values of the integrated Gaunt factor $g = P_{NR}^{ei}/P_{Kr}^{ei}$ [7], where $P_{Kr}^{ei}$ is the Kramers power of the electron-ion bremsstrahlung. For $T_e \ll 10$ eV $g = 1$; for $T_e \gg 10$ eV $g = 2\sqrt{3}/\pi \approx 1.1$ which corresponds to the non relativistic quantum Born approximation. At low $T_e$ (1–10 eV) Elwert approximation of the Gaunt factor for hydrogen plasma is $g_{Elw} \approx 1.5$ [7]. This value can be regarded as an upper bound. In accordance with the foregoing $\delta_B$ can be approximated by the following equation (also see Fig. 2):

$$\delta_B \approx \frac{Z_{eff}^3}{Z_{eff}^2}\left\{0.39\left[1-\exp(-0.008/\sqrt{\theta})\right]-0.49\exp(-505\sqrt{\theta})\right\}. \tag{4}$$

Electron-electron bremsstarhlung power is

$$P^{ee} = \frac{1}{2}n_e^2 \iiint \omega \frac{d\sigma^{ee}}{d\omega} u(\mathbf{p}_1,\mathbf{p}_2) f(\mathbf{p}_1) f(\mathbf{p}_2) d^3p_1 d^3p_2 d\omega, \tag{5}$$

where $d\sigma^{ee}$ is differential cross section of the process, $u(\mathbf{p}_1,\mathbf{p}_2)$ is the relative velocity of the electrons, the indices "1" and "2" denote the first and second colliding electrons, the factor 1/2 takes into account the identity of the electrons.

Integration of Eq. (5) with the non relativistic Born cross section $d\sigma^{ee}$ [8] and Maxwell distribution function leads to the expression [6, 9]

$$P_{NR}^{ee} = 4C_F \pi^{-1/2} C_b n_e^2 \theta^{3/2},$$

where $C_F = (5/9)(44-3\pi^2) \approx 8$.

In non relativistic limit $P_{NR}^{ee}/P_{NR}^{ei} \approx \frac{3}{\sqrt{2}} Z_{eff}^{-2} \theta \ll 1$. For $T_e < 100$ keV and $Z_{eff}^2 \sim 2$ accuracy of calculation of the electron-electron bremsstrahlung affects on the result less than the accuracy of calculating the electron-ion radiation. So one can not take into account the correction to the Born approximation. Evaluation of such a correction showed that for $T_e = 10$ and 20 keV power electron-electron emission in the Born approximation is overestimated by approximately 5 and 4% respectively [9].

Relativistic cross section $d\sigma^{ee}$ was obtained by Haug [10], who also suggested an approximate expression for the radiation power in the form of a double integral [11] with an accuracy of about 3%. By the same methodology to obtain an expression for electron-electron bremsstrahlung power was presented as a single integral [12]

$$P^{ee} = \frac{32 C_b n_e^2}{\theta [K_2(1/\theta)]^2} \int_1^\infty K_2\left(\frac{2x}{\theta}\right)\left[1-\frac{4}{3}\frac{\sqrt{x^2-1}}{x}+\frac{2}{3}(3-x^{-2})\ln(x+\sqrt{x^2-1})\right] \times$$
$$\times (x^2-1)^2 x dx. \tag{6}$$

Note that the integral expressions (Eqs. (3) and (6)) although derived using some approximations of the interim results, provide a calculation of the total bremsstrahlung emission from the plasma with high accuracy for $T_e > 1$ keV. In the extreme relativistic case ($\theta \gg 1$) these formulas give numerically the same results as the extreme relativistic expressions for electron-ion [6] and electron-electron [13] bremsstrahlung:

$$P_{ER}^{ei} = 12 C_b n_e^2 Z_{eff}^2 \theta\left[\ln(2\theta)+\tfrac{3}{2}-C_E\right],$$
$$P_{ER}^{ee} = 24 C_b n_e^2 \theta\left[\ln(2\theta)+\tfrac{5}{4}-C_E\right],$$

where $C_E = 0.5772...$ is the Euler constant. According the numerical results extreme relativistic expressions become practically applicable for $\theta > 2$ for the electron-ion radiation and for $\theta > 1$ for the electron-electron radiation

Numerical results for Maxwellian plasma can be approximated by the following formulas:

$$P^{ei} \approx \tfrac{32}{3}\sqrt{\tfrac{2}{\pi}} C_b n_e^2 Z_{eff}^2 \sqrt{\theta}\left[0.68+0.32\exp(-4.4\theta)+2.07\theta+\delta_B(\theta)\right], \quad \theta \leq 2; \tag{7}$$

$$P^{ee} \approx 4C_F \pi^{-1/2} C_b n_e^2 \theta^{3/2}(1+0.64\theta+6.6\theta^2-22.6\theta^3+33.8\theta^4-24.7\theta^5+7.1\theta^6), \quad \theta \leq 1. \tag{8}$$

The errors of these approximations are less then 3% at $T_e < 100$ keV and less then 5% at

$T_e > 100$ keV.

Fig. 3 shows a comparison of a number approximating dependencies with data of numerical calculations for a range of $T_e < 100$ keV. Eqs. (7) and (8) correspond to the solid lines passing through the circles which mark the numerical results. Note that the same level of agreement is demonstrated for analogous formulas [14, 15]. Drawing curves based on Eqs. (7), (8), the integral formulas Eqs. (3), (6), and the formulas from Refs. 14, 15 are indistinguishable at present figure scale. The differences between these formulas are within a few percent.

A satisfactory agreement for the power of the relativistic electron-ion bremsstrahlung can be seen for formula suggested by Gould [5]. According this formula, $P^{ei} \approx 8.5 C_b n_e^2 Z_{eff}^2 \sqrt{\theta}(1 + 0.8\theta + 1.87\theta^2)$. Since the electron-electron emission was considered in Ref. 5 as the first order correction (order of θ) to the electron-ion emission. The corresponding expression coincides with non relativistic limit (lower dashed curve in Fig. 3).

Often the calculation of bremsstrahlung at relativistic temperatures used Dawson formula [16]. It was obtained as an approximation of the results and of interpolation between the non relativistic and extreme relativistic limit cases performed by Maxon [17]. According Dawson formula $P^{ei} \approx 8.5 C_b n_e^2 Z_{eff}^2 \sqrt{\theta}(1 + 2\theta)$ and $P^{ee} \approx 17 C_b n_e^2 \sqrt{\theta}(1 + 2\theta)[1 - 1/(1 + \theta)]$. At $T_e \approx$ 50–80 keV Dawson formula gives the values of bremsstrahlung emission power 10–15 % higher then numerical results.

*This work was supported by the Ministry of Education and Science of Russian Federation in framework of Presidential Program, grant MK-1811.2010.8.*

**References**

1. Chirkov A.Yu., Khvesyuk V.I., Ryzhkov S.V., in V Int. Symp. on Radiation Plasma Dynamics, Moscow, 2000. P. 95–96.
2. Bekefi G. Radiation Processes in Plasmas. New York, Wiley, 1966.
3. Akhiezer A.I., Berestetskii V.B. Quantum electrodynamics. Moscow, Nauka, 1981.
4. Lee C.M., Kissel L., Pratt R.H., Tseng H.K. // Phys. Rev. A. 1976. V. 13. P. 1714–1727.
5. Gould R.J. // Astrophys. J. 1980. V. 238. P. 1026–1033.
6. Stickforth J. // Z. Physik. 1961. V. 164. P. 1–20.
7. Greene J. // Astrophys. J. 1959. V. 130. P. 693–701.
8. Fedyushin B.K. // JETP. 1952. V. 22, No. 2. P. 140–142.
9. Maxon S.M., Corman E.G. // Phys. Rev. 1967. V. 163, No. 1. P. 156–162.
10. Haug E. // Z. Naturforsch. 1975. V. 30a. P. 1099–1113.
11. Haug E. // Z. Naturforsch. 1975. V. 30a. P. 1546–1552.
12. Kukushkin A.B., Kogan V.I. // Fiz. Plazmy. 1979. V. 5. P. 1264–1268.
13. Alexanian M. // Phys. Rev. 1968. V. 165, No. 1. P. 253–257.
14. Svensson R. // Astrophys. J. 1982. V. 258. P. 335–348.
15. Khvesyuk V.I., Chirkov A.Yu. // Tech. Phys. Letters. 2000. V. 26, No. 11. P. 964–966.
16. Dawson J.M. in Fusion, Ed. E. Teller. New York, Academic Press, 1981. Vol. 1. Part B. Chapt. 16.
17. Maxon S. // Phys. Rev. A 1972. V. 5, No. 4. P. 1630–1633.

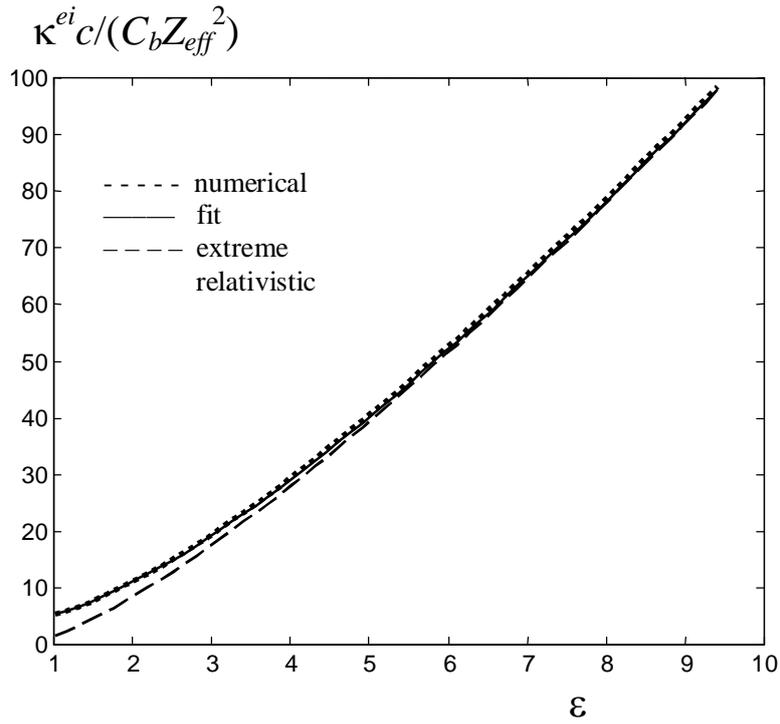

Fig. 1. The electron energy losses during slow-down on the ions:
- - - - - - calculation in Born approximation; ──────── approximation by Eq. (2);
– – – – – extreme relativistic limit

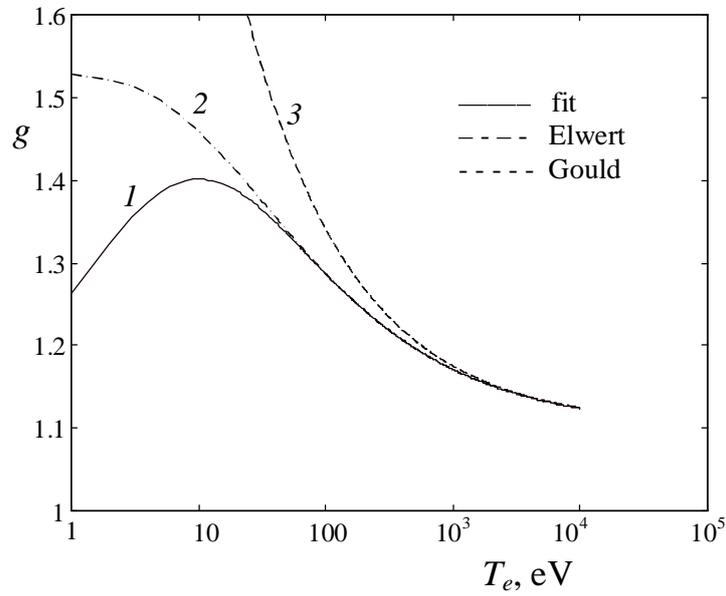

Fig. 2. Gaunt factors at low temperatures for various approximations of $\delta_B$:
*1* – Eq. (4) providing $g \to 1$ at $T_e \to 0$; *2* – $g \to g_{Elw}$ at $T_e \to 0$; *3* – by Gould [5]

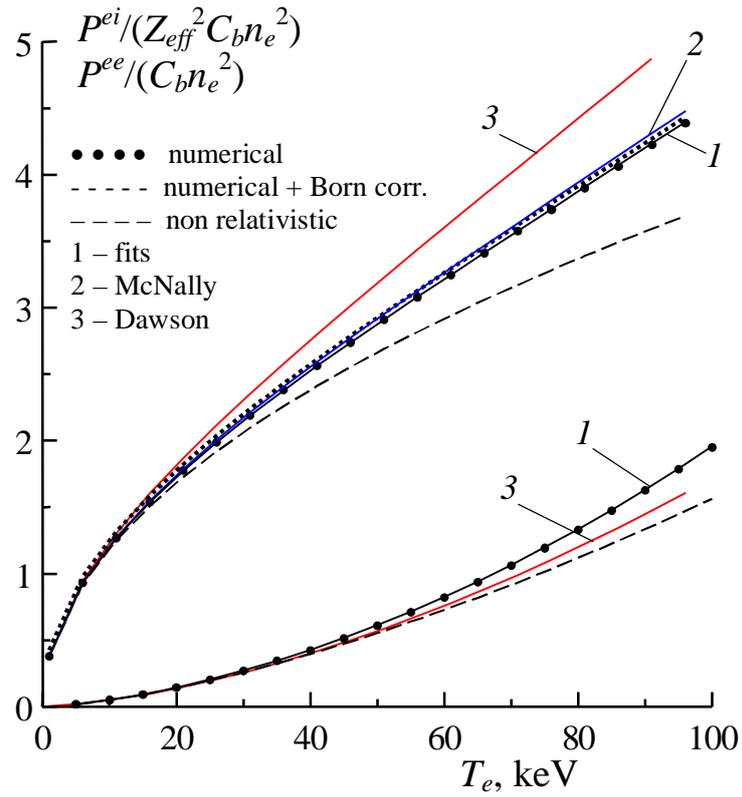

Fig. 3. Electron bremsstrahlung emission per unit plasma volume on ions (upper curves) and electrons (lower curves) at $T_e$ up to 100 keV:
● ● ● ● numerical calculation in the Born approximation;
- - - - - - numerical calculation with the Gould correction;
——— non relativistic expressions;
*1* – Eqs. (7) and (8) without correction to the Born approximation;
*2* – according Ref. 5 without correction to the Born approximation;
*3* – Dawson formula